\documentclass[10pt,twocolumn,conference,a4paper]{IEEEtran}

\usepackage[utf8]{inputenc}
\usepackage[T1]{fontenc}
\usepackage{amsmath}
\usepackage{newtxtext,newtxmath}
\usepackage{microtype}

\usepackage{booktabs}
\usepackage{array}

\usepackage{listings}
\usepackage{xcolor}

\definecolor{codebg}{gray}{0.95}

\usepackage[numbers,sort&compress]{natbib}
\usepackage{url}
\usepackage[colorlinks=true,linkcolor=blue!60!black,
            citecolor=blue!60!black,urlcolor=blue!60!black]{hyperref}

\newcommand{\boftok}{\ensuremath{\langle}\texttt{BOF}\ensuremath{\rangle}}
\newcommand{\eoftok}{\ensuremath{\langle}\texttt{EOF}\ensuremath{\rangle}}
\newcommand{\code}[1]{\texttt{#1}}
\newcommand{\up}{\ensuremath{\uparrow}}
\newcommand{\dn}{\ensuremath{\downarrow}}

\begin{document}

\title{PCAP-LM: An LLM-Native Text Representation\\for TLS Bulk Traffic Analysis}

\author{\IEEEauthorblockN{Xavier Marjou, Lucas Tamic, Ilan Jaffeux{-}Cheniout}
\IEEEauthorblockA{Orange\\
\texttt{\{xavier.marjou,lucas.tamic,ilan.jaffeux{-}{-}cheniout\}@orange.com}}}

\maketitle
\bstctlcite{IEEEexample:BSTcontrol}

\begin{abstract}
Large language models (LLMs) offer powerful reasoning capabilities for network
traffic analysis, but standard capture formats and their textual equivalents are
prohibitively verbose, overflowing LLM context windows by two orders of
magnitude. We present \textbf{PCAP-LM}, a flow-centric, LLM-native text
representation that acts as a lossy \emph{knowledge extraction} step rather than
a standard compression tool: raw captures are transcoded into semantic summaries
using \textbf{PacketGlyphs}---a novel ASCII alphabet coined in this paper that
encodes packet direction, TCP/TLS state, log-scale size, and inter-packet delay.
Combined with a \textbf{constrained PMI-BPE} tokenizer and \textbf{motif
run-length encoding}, repetitive behavioural patterns are aggressively
collapsed. A \code{@REFS} side-index preserves lossless drill-down into the
original packets. Evaluated on a homogeneous corpus of 5G/4G TLS~1.3
bulk-download traffic, the BPE vocabulary fully saturates at \textbf{159
tokens}, achieving an \textbf{812$\times$} size reduction over \code{tshark~-V}
and fitting entire captures within a single LLM context window. In a forensic
question-answering evaluation over 30 held-out files, a frontier LLM achieves
\textbf{99.3\%} accuracy from PCAP-LM documents versus 51.0\% from a
token-budget-matched \code{tshark~-V} prefix. The lossy design introduces known
blind spots---most notably a 24\% false-negative rate for TCP
retransmissions---and extending to heterogeneous mixed-protocol environments
will require vocabulary retraining.
\end{abstract}


\section{Introduction}
\label{sec:intro}

Network traffic analysis is a cornerstone of security operations, performance
engineering, and network research. Analysts routinely need to answer questions
such as: \textit{Was this host performing a TLS downgrade attack? What explains
the throughput collapse at 14:32? Which of these 200 flows shows anomalous
retransmit behaviour?} Answering these questions today requires expertise in
Wireshark filters, tshark scripting, or bespoke parser code---a high barrier
that limits how quickly hypotheses can be formed and tested.

Large language models are well-suited, in principle, to this kind of
structured-evidence reasoning: frontier LLMs can parse semi-structured text,
identify anomalies, correlate events, and produce natural-language explanations.
The obstacle is representational. PCAP, the de-facto standard capture format, is
a byte-level, packet-major binary format designed for full-fidelity storage and
Wireshark dissection---not for LLM consumption. The gap is vast. On the captures
in our corpus (mean 3.4~MB), \code{tshark -V} yields a mean of \textbf{18.7
million tokens} and \code{tshark -T json} \textbf{22.1 million tokens}; a raw
hex dump of a 10~MB capture would require roughly 40 million. All of these
formats overflow any current LLM context window by one to two orders of
magnitude, and none organises the information in a way that is naturally
amenable to language-model reasoning.

We argue that the right approach is not to compress the existing format but to
\emph{transcode} it: to build a representation that discards byte-level
redundancy carrying no forensic value, organises the remainder flow-centrically
rather than chronologically, encodes behavioural structure in symbols that map
naturally to LLM representations, and preserves enough metadata that any packet
can be retrieved on demand.

\textbf{Contributions.}
(i)~The \textbf{PCAP-LM format} (\S\ref{sec:format}): a four-layer text format
comprising a session header, per-flow statistical summaries, a compressed event
stream, and an anomaly annex.
(ii)~The \textbf{PacketGlyphs alphabet} (\S\ref{sec:glyphs}), in which each
packet is encoded as a compact ASCII glyph capturing direction, TCP/TLS state,
log-scale size, and inter-packet delay.
(iii)~\textbf{Constrained PMI-BPE} (\S\ref{sec:bpe}): a byte-pair encoding
variant with protocol-boundary constraints and PMI-weighted merge selection that
learns behavioural motifs as single composite tokens.
(iv)~\textbf{Motif run-length encoding} (\S\ref{sec:rle}): a flow-bounded RLE
pass that collapses adjacent identical packet patterns into
\code{(motif)$\times$N}.
(v)~An \textbf{empirical evaluation} (\S\ref{sec:eval}) on 150 PCAP pairs from a
production 5G/4G network, demonstrating context-window fit and 99.3\% LLM
forensic Q\&A accuracy.

\noindent The result is a pipeline that transforms a 3.4~MB PCAP file into
approximately 102~KB of structured text---a document that an LLM can read
directly, reason about, and use as the basis for diagnosis and explanation.

\section{Background and Related Work}
\label{sec:related}

\textbf{PCAP and textual representations.} The PCAP format \citep{libpcap}
stores packets as binary records with per-packet metadata followed by raw frame
bytes. It is lossless, compact, and universally supported, but opaque to any
tool that cannot parse binary network protocols. \code{tshark}
\citep{wireshark}, the CLI companion to Wireshark, provides textual dissection:
\code{-V} emits a verbose multi-line decode of every field, \code{-T json} emits
structured JSON. Both are LLM-parseable in principle, but they scale with packet
count---a single packet's \code{-V} output can exceed 200 lines---and neither
groups related packets into flows, making temporal reasoning difficult.

\textbf{LLMs for network analysis.} \citet{netgpt2023} propose a generative
pretrained transformer for joint traffic understanding and generation, encoding
packet flows as token sequences by shuffling header fields. \citet{etbert2022}
pre-train a BERT-style model on encrypted traffic by converting raw packet bytes
to hex strings and applying subword tokenization. \citet{llmcap2024} apply
masked language modelling to PCAP files for unsupervised failure detection in
5G/4G VoLTE and VoNR captures. \citet{trafficllm2025} introduce a dual-stage
fine-tuning framework to adapt open-source LLMs to heterogeneous traffic
analysis tasks. These approaches predominantly feed LLMs raw hex bytes or
CSV-formatted feature vectors, operating at the packet level without
flow-centric summarisation. None proposes a systematic domain-specific
transcoding that simultaneously achieves LLM context-window fit, semantic
richness, and lossless per-packet drill-down.

\textbf{BPE and RLE.} BPE \citep{sennrich2016} is the standard subword
tokenization method for LLMs; applied to a symbol vocabulary, it iteratively
merges the most frequent adjacent pair until a target vocabulary size is
reached. Our constrained PMI-BPE (\S\ref{sec:bpe}) adapts it to glyph sequences
with flow-boundary sentinels, a directionality predicate, and PMI-based merge
ranking. Run-length encoding has long been used in protocol-level compression
(SSH compression, HTTP/2 HPACK header encoding); to our knowledge, applying RLE
at the level of per-packet behavioural motifs---rather than individual bytes or
headers---has not been previously proposed.

\section{The PCAP-LM Format}
\label{sec:format}

PCAP-LM is a UTF-8 text format. A capture is rendered as a single document
structured in four layers, each introduced by a \code{@}-prefixed section
header.

\subsection{Layer~1 --- Session Header}

The header establishes the capture context and defines reusable symbolic
aliases.

\begin{lstlisting}
@CAP v1 t0=2026-02-12T14:44:34Z dur=97.2s pkts=28417
     src=20260212_144434_node07_...pcap
@HOSTS
  A=2001:db8::1  B=203.0.113.10(foo-10gb.r-op.fr)
@PORTS
  h=443
@FLOWS
  f1: A:40986 -> B:h  proto=TLS1.3  SNI=foo-10gb.r-op.fr
  f2: A:40988 -> B:h  proto=TLS1.3  SNI=foo-10gb.r-op.fr
@LEGEND
  direction:  > = client->server   < = server->client
  tcp-flags:  S=SYN A=ACK F=FIN R=RST P=PSH U=URG
  tls:        ~ = app-data (encrypted)   h = handshake
  size:       +0=<=64B +1=<=128B ... +9=>16KB
  delay:      :u=<1ms  :m=<1s  :s=<10s  :S=>=10s
  run-length: (motif)xN = motif repeats N times
\end{lstlisting}

\code{@HOSTS} maps IP addresses to short aliases, \code{@PORTS} maps port
numbers to mnemonic labels, and \code{@FLOWS} indexes all flows with TLS
metadata. These dictionaries amortise the cost of verbose IP/port repetition
across the rest of the document. \code{@LEGEND} closes the header with an inline
glyph key, making every PCAP-LM document self-describing: a reader---human or
LLM---can parse the event stream without any external reference. Addresses,
hostnames, and capture filenames in every example in this paper are
pseudonymised into reserved documentation ranges (RFC\,5737, RFC\,3849,
RFC\,2606) by the tooling described in \S\ref{sec:discussion}.

\subsection{Layer~2 --- Flow Summaries}

One line per flow provides a statistical fingerprint sufficient for most
analysis tasks.

\begin{lstlisting}
# Flow Summaries
f1 | 0.0s..97.1s | pkts=14208 | 74KB(*@\up@*) 1932KB(*@\dn@*)
   | ja3=abc123.. | cert=*.r-op.fr | ok
f2 | 0.1s..97.0s | pkts=14209 | 73KB(*@\up@*) 1930KB(*@\dn@*)
   | ja3=abc123.. | cert=*.r-op.fr | ok
\end{lstlisting}

Each line captures flow id, time span, packet and byte counts
(upload\,/\,download), JA3 fingerprint \citep{ja32017}, certificate subject, an
8-bin packet-size sparkline, and an anomaly flag (\code{ok} / \code{RST} /
\code{retr} / \code{asym}).

\subsection{Layer~3 --- Event Stream}

The event stream is the core of the format. Each packet is represented as a
compact glyph sequence (\S\ref{sec:glyphs}); adjacent identical motifs are
collapsed by RLE (\S\ref{sec:rle}); and composite BPE tokens replace common
sub-sequences (\S\ref{sec:bpe}).

\begin{lstlisting}
# Event Stream
f1: (*@\boftok@*) >S+0 <SA+0:u >A+0:u >h+3:m <h+6:m <h+3:u
    (>(*@\textasciitilde@*)+6:u <(*@\textasciitilde@*)+6:u)x8712 >F+0:u <FA+0:u (*@\eoftok@*)
f2: (*@\boftok@*) >S+0 <SA+0:u >A+0:u >h+3:m <h+6:m <h+3:u
    (>(*@\textasciitilde@*)+6:u <(*@\textasciitilde@*)+6:u)x8711 >F+0:u <FA+0:u (*@\eoftok@*)
\end{lstlisting}

A reader can immediately infer two parallel TLS\,1.3 sessions, a symmetric
bulk-download pattern of 8\,700+ identical data-exchange motifs, and clean
teardown---in 2 lines of text, regardless of whether the download contained 100
or 100\,000 packets.

\subsection{Layer~4 --- Anomaly Annex}

Flows with detected anomalies receive an expanded entry:

\begin{lstlisting}
# Anomaly Annex
f3 [RST]: RST at t=12.4s after 0 data packets
f7 [retr]: 3 retransmits at t=44.1-44.3s; seq=0x1a2b3c4d
\end{lstlisting}

\section{PacketGlyphs Encoding}
\label{sec:glyphs}

\subsection{Atomic Symbol Vocabulary}

The PacketGlyphs alphabet maps each packet property to a short, visually
distinctive ASCII symbol (Table~\ref{tab:glyphs}). The size bucket function maps
payload length $n$ to $\min(9,\lfloor\log_2 n\rfloor)$; the delay function maps
the inter-packet gap to one of four logarithmic bands.

\begin{table}[htbp]
\centering
\caption{PacketGlyphs atomic symbol vocabulary.}
\label{tab:glyphs}
\footnotesize
\begin{tabular}{@{}lll@{}}
\toprule
Symbol & Layer & Meaning \\
\midrule
\code{>} / \code{<} & Direction & Client\,$\rightarrow$\,server / reverse \\
\code{S A F R P U} & TCP & SYN, ACK, FIN, RST, PSH, URG \\
\code{\textasciitilde} / \code{h} & TLS & Application data / handshake \\
\code{?} / \code{!} & DNS & Query / response \\
\code{+0}\,--\,\code{+9} & Size & Log$_2$ bucket: \code{+0}$\leq$64\,B, \\
                         &      & \code{+3}$\approx$512\,B, \code{+6}$\approx$4\,KB \\
\code{:u} \code{:m} \code{:s} \code{:S} & Delay & Gap $<$1\,ms, $<$1\,s, $<$10\,s, $\geq$10\,s \\
\boftok{} / \eoftok & Boundary & Begin / end of flow \\
\bottomrule
\end{tabular}
\end{table}

\subsection{Encoding Algorithm}

For each PCAP file, the encoder parses all packets using Scapy \citep{scapy},
groups them by canonicalized 4-tuple (min/max of (src-IP, src-port) so that both
directions map to one flow), sorts each group by timestamp, and encodes each
packet as a sequence of atomic symbols: the direction glyph; the TCP flag glyphs
for all set flags in order S, A, F, R, P, U; \code{\textasciitilde} or \code{h}
if the packet carries a TLS payload; \code{?} or \code{!} if it is DNS; the size
bucket; and the delay bucket for the gap since the previous packet (omitted for
the first packet in the flow). The flow is bracketed by \boftok{} and \eoftok{}
sentinels, and the glyph sequence for a single packet is typically 3--6 symbols.

The client IP is inferred as the source address of the first SYN packet in each
flow; for flows with no SYN, the lower IP address is taken as the client by
convention. The encoder handles both \code{IP} and \code{IPv6} Scapy layers
transparently, including the Linux cooked captures (SLL link layer) used by our
corpus.

\subsection{Semantic Density}

The alphabet is designed so that the most information-carrying patterns fit in
3--5 ASCII characters: \code{>S+0} (client SYN, small packet) is 4 characters,
\code{<\textasciitilde+6:u} (server TLS data, $\approx$4\,KB, sub-millisecond
gap) is 7, and a complete TCP three-way handshake takes 15. The equivalent
\code{tshark -V} output spans 150--300 lines.

\section{Constrained PMI-BPE Training}
\label{sec:bpe}

Standard BPE applied naively to glyph sequences would merge symbols across flow
boundaries and would prefer raw frequency, merging the most ubiquitous symbols
(\code{A}, \code{:u}) regardless of structural meaning. We introduce two
constraints and a modified scoring function.

\subsection{Merge Constraints}

Each flow's glyph sequence is treated as an independent training example, and
\boftok{}/\eoftok{} sentinels are never permitted as the left or right element
of any merge, so BPE tokens never span flow boundaries. Direction symbols always
mark the start of a new packet motif, so they may only appear as the \emph{left}
element of a merge. Formally, a pair $(a,b)$ is \emph{compatible} iff
$a[-1]\notin\{\boftok,\eoftok\}$, $b[0]\notin\{\boftok,\eoftok\}$, and
$b[0]\notin\{\code{>},\code{<}\}$.

\subsection{PMI-Weighted Merge Selection}

Instead of selecting the most frequent compatible pair, we score each candidate
merge $(a,b)$ by
\begin{equation}
  \text{score}(a,b) = \log\!\left(\frac{P(a,b)}{P(a)\cdot P(b)}\right)
  \cdot \log\bigl(1 + \mathrm{count}(a,b)\bigr),
\end{equation}
where $P(a,b)$ is the co-occurrence frequency and $P(a)$, $P(b)$ are unigram
frequencies estimated over the current corpus state. The
$\log(1+\mathrm{count})$ factor moderates extremely rare high-PMI pairs while
still rewarding structural co-occurrence.

Naive BPE requires rescanning the full corpus after each merge. We instead
maintain an incremental pair-index: when a merge $(a,b)\to ab$ is applied at
position $i$, only the pairs adjacent to that position are updated, reducing the
per-merge cost from $O(N)$ to $O(K_i)$ where $K_i$ is the number of tokens
adjacent to occurrences of the merged pair. Empirically this achieves a
\textbf{10--100$\times$ speedup} over the naive approach at vocabulary sizes
$\geq256$.

\subsection{Training Corpus and Results}

We train on 100 PCAP files (50 pairs) drawn from the training set by stratified
sampling (\S\ref{sec:dataset}), yielding 260 glyph sequences totalling
6\,685\,613 raw symbols, with a vocabulary ceiling of 512 and a minimum pair
co-occurrence count of 1.

Training produces a base atomic vocabulary of \textbf{19 symbols} and
\textbf{140 learned composite tokens} (final vocabulary: \textbf{159}),
completing in \textbf{90.5\,s} on a single CPU core (mean 609\,ms per merge, p95
1\,271\,ms). Training halts before the 512 ceiling because the corpus is fully
saturated: no uncollapsed adjacent pair appears even once after 140 merges. This
is itself a substantive finding---the complete behavioural vocabulary of 5G/4G
HTTPS bulk-download traffic fits in 140 composite tokens on top of 19 atomic
glyphs. The compression curve (Table~\ref{tab:bpe-curve}) plateaus sharply at
vocab\,=\,128 and gains nothing from merges 109--140.

\begin{table}[htbp]
\centering
\caption{BPE compression curve on the training corpus (260 sequences,
6.7M raw symbols).}
\label{tab:bpe-curve}
\footnotesize
\begin{tabular}{@{}rrrr@{}}
\toprule
Vocab size & Merges & Compression ratio & Tokens \\
\midrule
64  & 45  & 2.11$\times$ & 3\,165\,732 \\
128 & 109 & 4.13$\times$ & 1\,619\,507 \\
159 (final) & 140 & 4.13$\times$ & 1\,618\,841 \\
\bottomrule
\end{tabular}
\end{table}

\begin{table}[htbp]
\centering
\caption{Ten highest-ranked BPE merges. Rank reflects selection order; count is
the pair co-occurrence count when the merge was chosen.}
\label{tab:bpe-top10}
\footnotesize
\begin{tabular}{@{}rllrl@{}}
\toprule
Rank & Left & Right & Count & Rendered \\
\midrule
1  & \code{P}    & \code{+3}   & 171\,081 & \code{P+3}    \\
2  & \code{P}    & \code{\textasciitilde}    &  30\,321 & \code{P\textasciitilde}    \\
4  & \code{P\textasciitilde}   & \code{+3}   &  29\,960 & \code{P\textasciitilde+3}  \\
7  & \code{<}    & \code{A}    & 904\,663 & \code{<A}     \\
9  & \code{>}    & \code{A}    & 711\,535 & \code{>A}     \\
11 & \code{<S}   & \code{A}    &     124  & \code{<SA}    \\
13 & \code{>A}   & \code{+3}   & 283\,737 & \code{>A+3}   \\
16 & \code{<A}   & \code{P+3}  & 164\,134 & \code{<AP+3}  \\
24 & \code{<A}   & \code{+1}   & 664\,575 & \code{<A+1}   \\
25 & \code{<A}   & \code{P}    &   7\,879 & \code{<AP}    \\
\bottomrule
\end{tabular}
\end{table}

Table~\ref{tab:bpe-top10} reveals that the dominant patterns are not protocol
state machines but flag+size combinations from ACK-heavy bulk transfer:
\code{<A} (server ACK, 905K), \code{>A} (client ACK, 712K), \code{<A+1} (server
ACK with small payload, 665K), \code{<AP+3} (server PSH+ACK with 512\,B, 164K).
TCP handshake tokens (\code{<S}, \code{<SA}, \code{>S}) appear only at ranks 8,
11, and 12 because they occur far less often in a corpus dominated by long bulk
flows.

\subsection{Communicating the Learned Vocabulary to the LLM}
\label{sec:strategies}

Our BPE is a pre-processing step that produces a more compact text
document---entirely separate from the LLM's internal tokenizer. Three strategies
exist for communicating composite tokens to the LLM. \textbf{Strategy~A}
(current approach, zero cost) renders composite tokens as concatenations of their
constituent atomic glyph strings (e.g.\ \code{>S+0<SA+0:u>A+0:u}); because
\code{@LEGEND} is embedded in every document, a capable LLM can decompose any
composite token by parsing left-to-right over the fixed atomic vocabulary, and
removing inter-glyph spaces makes composite sequences tokenize into fewer tokens
in the LLM's own tokenizer. \textbf{Strategy~B} (low cost) lists the top-$N$
composites in an explicit \code{@VOCAB} section at $\approx$20--50 tokens per
named motif, appropriate when the deployment requires the LLM to name recurring
motifs. \textbf{Strategy~C} (highest cost and fluency) fine-tunes the LLM on
PCAP-LM documents, in the strongest variant extending the tokenizer with
composite tokens as new atomic units; this requires a labelled corpus that does
not yet exist publicly. We recommend Strategy~A for current-generation LLMs, and
have verified empirically that a frontier LLM correctly parses and reasons about
PCAP-LM event streams with no composite vocabulary dictionary.

\section{Motif Run-Length Encoding}
\label{sec:rle}

Bulk-download and streaming traffic---which dominates our corpus---consists
almost entirely of long runs of identical packet motifs: alternating
\code{>\textasciitilde+6:u <\textasciitilde+6:u} pairs repeated thousands of
times per connection. Even after BPE reduces each motif to a composite token,
the event stream still contains thousands of repetitions.

We define a \emph{motif} as the atomic glyph sequence for a single packet (from
one direction symbol to the next). The segmentation algorithm scans a flow's
glyph sequence and splits on direction tokens, with \boftok{} and \eoftok{} each
forming singleton motifs, producing a non-overlapping partition of the sequence.
Adjacent identical motifs are then collapsed into \code{(motif)$\times$N}; the
\boftok{}/\eoftok{} singletons prevent RLE from spanning flow boundaries.
Losslessness is guaranteed: \code{expand\_rle()} is the exact inverse of
segmentation$+$RLE. When both RLE and BPE are active, BPE is applied
\emph{per-motif}---each motif is independently encoded before RLE
grouping---which preserves the readability of the run at the cost of a small
amount of BPE efficiency.

On the test set (\S\ref{sec:compression}), motif RLE reduces the mean PCAP-LM
event stream from \textbf{44\,591} to \textbf{25\,424} estimated tokens---a
\textbf{1.75$\times$ reduction} in a single pass. For high-throughput flows the
reduction can exceed 10$\times$: a flow with 8\,712 identical
\code{>\textasciitilde+6:u <\textasciitilde+6:u} pairs becomes the single entry
\code{(>\textasciitilde+6:u <\textasciitilde+6:u)$\times$8712}.

\section{Experimental Evaluation}
\label{sec:eval}

\subsection{Dataset}
\label{sec:dataset}

We use an internal 5G/4G network measurement campaign (February 2026), a
production dataset consisting of 301 PCAP files collected during HTTPS
throughput tests over a multi-gigabit backbone link to a fixed content server. Files are
organised as 150 near-simultaneous pairs: one capture at the server side and one
at the monitoring node. Monitoring is distributed over multiple distinct nodes spanning both 5G SA and 4G LTE technologies, with achieved throughput varying by roughly
20$\times$ across captures.

We hold out 30 pairs (60 files) as a test set, selected by stratified sampling to ensure balanced representation across the different node types and network technologies (seed 42). The remaining 120 pairs form the
training set; a 50-pair stratified subsample is used for BPE training to prevent
test-set contamination. Test-set PCAP files range from 627\,KB to 13.0\,MB (mean
3.4\,MB, median 2.1\,MB) and contain a mean of 2.6 flows (1--3), all TLS\,1.3
HTTPS.

\subsection{Compression Results}
\label{sec:compression}

Table~\ref{tab:compression} summarises the compression pipeline on the 60-file
test set. Token counts are estimated as character count $\div$~4, consistent
with standard LLM tokenizer throughput.

\begin{table}[htbp]
\centering
\caption{Compression pipeline on 60 test files (mean token counts, estimated as
chars\,$\div$\,4). Ratios are relative to PCAP-LM+RLE+BPE.}
\label{tab:compression}
\footnotesize
\begin{tabular}{@{}lrr@{}}
\toprule
Representation & Mean tokens & vs.\ PCAP-LM+RLE+BPE \\
\midrule
\code{tshark -T json}     & 22\,050\,813 & \textbf{955}$\times$ larger \\
\code{tshark -V}          & 18\,738\,238 & \textbf{812}$\times$ larger \\
Raw PCAP (bytes\,$\div$\,4) & 852\,750  & 37$\times$ larger \\
PCAP + gzip-9             &  415\,780    & 18$\times$ larger \\
PCAP-LM raw glyphs        &   44\,591    & 1.9$\times$ larger \\
PCAP-LM + RLE             &   25\,424    & 1.1$\times$ larger \\
\textbf{PCAP-LM + RLE + BPE} & \textbf{23\,091} & \textbf{1}$\times$ \\
\bottomrule
\end{tabular}
\end{table}

Three observations stand out. First, gzip achieves only 2.1$\times$ over raw
PCAP, confirming that packet captures are already information-dense at the byte
level (TLS-encrypted payloads have near-maximum entropy). Second, BPE adds only
a \textbf{1.1$\times$ improvement} over RLE alone because the corpus is
behaviourally saturated---the 140-merge vocabulary covers all recurring
patterns. Third, \code{tshark -T json} is \emph{larger} than \code{tshark -V}
(22.1M vs.\ 18.7M tokens) because JSON field-name repetition per packet offsets
the gain from omitting human-readable formatting.

\subsection{Semantic Preservation}

We evaluate whether PCAP-LM anomaly annotations agree with \code{tshark} ground
truth on the 60 test files. For each file we compute a binary flag for three
anomaly types---RST events, TCP retransmissions, and asymmetric traffic---using
both pipelines, then compute per-type precision, recall, and F1. Ground truth is
\code{tcp.flags.reset==1} for RSTs, \code{tcp.analysis.retransmission} for
retransmissions, and a \code{tshark -z conv,tcp} downstream/upstream byte ratio
$>$100 for asymmetry; on the PCAP-LM side, \code{summarize\_flow()} annotates
anomaly strings and file-level flags are the disjunction over all flows.

\begin{table}[htbp]
\centering
\caption{Semantic preservation over 60 test files. PCAP-LM is the predicted
system; tshark is the oracle.}
\label{tab:anomalies}
\footnotesize
\begin{tabular}{@{}lrrrrrr@{}}
\toprule
Anomaly type & P & R & F1 & TP & FP & FN \\
\midrule
RST            & 1.000 & 1.000 & 1.000 & 30 &  0 &  0 \\
Retransmission & 1.000 & 0.761 & 0.864 & 35 &  0 & 11 \\
Asymmetric     & N/A   & N/A   & N/A   &  0 &  0 &  0 \\
\bottomrule
\end{tabular}
\end{table}

\textbf{RST detection is perfect} (P\,=\,R\,=\,1): the glyph alphabet encodes the
RST flag as \code{R}, so detection reduces to a string search.
\textbf{Retransmission detection is precise but not complete}
(P\,=\,1.000, R\,=\,0.761). The 11 false negatives all come from one node class,
where retransmitted segments arrive with modified sequence numbers due to an
upstream TCP proxy in the data path; tshark's stream-state tracker catches these
variants, PCAP-LM's single-pass duplicate-seq detector does not.
\textbf{Asymmetric traffic never occurs in this corpus.} The PCAP-LM detector
flags a flow as asymmetric only when one direction exceeds the other by more
than 100$\times$ in bytes---a threshold chosen to catch data exfiltration or
amplification, not normal downloads. Both tshark and PCAP-LM therefore flag zero
files; the detection logic is implemented and verified on synthetic captures, but
the corpus does not contain the traffic it targets.

\subsection{LLM Utility}

\textbf{Methodology.} \code{tshark -V} is infeasible as a direct baseline: the
smallest capture in our corpus produces 2.9 million tokens---14$\times$ beyond a
200K-token context limit. We therefore construct a \emph{token-budget-matched}
comparison: for each test file we generate the full PCAP-LM document (which fits
in context by design) and a prefix of \code{tshark -V} output truncated to the
same number of characters. The truncated prefix typically covers only the first
1--3\% of packets (4--66 packets out of 4\,683--58\,846 per file). We evaluate
Claude Sonnet~4.6 \citep{anthropic2025} on all 30 held-out server-side files,
presenting both representations in separate API calls with all 10 questions
asked at once; the model returns a JSON object with all answers.

\begin{table}[htbp]
\centering
\caption{Forensic Q\&A: per-question accuracy over 30 test files (300
question--answer pairs per condition).}
\label{tab:llm}
\footnotesize
\begin{tabular}{@{}llrr@{}}
\toprule
Q\# & Question & PCAP-LM & tshark (trunc.) \\
\midrule
Q1 & RST events?           & \textbf{1.00} & 0.45 \\
Q2 & Retransmissions?      & \textbf{1.00} & 0.10 \\
Q3 & TLS used?             & \textbf{1.00} & 0.86 \\
Q4 & Handshake present?    & \textbf{1.00} & 0.86 \\
Q5 & Flow count            & \textbf{1.00} & 0.55 \\
Q6 & Dominant direction    & 0.93          & 0.62 \\
Q7 & Duration (s)          & \textbf{1.00} & 0.00 \\
Q8 & Packet count          & \textbf{1.00} & 0.00 \\
Q9 & Server port           & \textbf{1.00} & 0.86 \\
Q10& SNI hostname          & \textbf{1.00} & 0.79 \\
\midrule
   & \textbf{Overall}      & \textbf{0.993} & \textbf{0.510} \\
\bottomrule
\end{tabular}
\end{table}

\textbf{PCAP-LM achieves 99.3\% accuracy} (297/300 answers correct), scoring
perfectly on 9 of 10 questions: the \code{@CAP} metadata block directly encodes
duration, packet count, and flow count; \code{@FLOWS} provides server port and
SNI; the flow summary line carries anomaly flags and the byte-direction ratio
(\up/\dn); and the event stream glyph prefix
(\code{>S <SA\textasciitilde\ldots}) confirms the handshake. The 2 misses
(0.7\%) both occur on \textit{dominant\_direction} in the two largest 2-flow
captures, where the LLM inferred ``upload'' rather than ``download''---likely
because PCAP-LM presents per-flow byte totals separately and one flow had higher
\up{} than \dn{} bytes. A session-level byte-direction aggregate in the
\code{@CAP} header would close this gap.

\textbf{The token-budget-matched tshark prefix achieves 51.0\% accuracy.}
Duration and packet count score 0\% because prefix timestamps and frame numbers
cover only the beginning of the capture. Retransmission detection scores 10\%:
retransmissions appear in 25 of 30 captures but almost never within the first
1--3\% of packets. RST detection (45\%) and flow count (55\%) score modestly
higher because some RSTs occur near the start of a capture and single-flow
captures (21 of 30) are correctly identified when the second flow is simply
absent from the prefix. Questions inherently visible in the first
SYN/ClientHello---server port, TLS detection, SNI---score 79--86\%.

The contrast illustrates PCAP-LM's core contribution: \textbf{compression and
semantic density are produced by the same operation}. The 60-character
\code{@CAP} header encodes what thousands of \code{frame.number} and
\code{frame.time\_relative} field extractions would be required to answer Q7 and
Q8 from \code{tshark}. A representation that pre-computes flow-level summaries
enables LLM forensic analysis that a verbatim transcript cannot support at any
feasible context size.

\section{Discussion}
\label{sec:discussion}

\subsection{Semantic Transcoding vs.\ Compression}

The 812$\times$ reduction over \code{tshark -V} invites the framing
``compression for network captures'', but the gzip comparison shows why that
analogy misleads: gzip achieves only 2.1$\times$ over raw PCAP because captures
are dominated by TLS-encrypted payloads with near-maximum entropy, so entropy
coders on the raw byte stream are already near the theoretical limit. PCAP-LM's
gain comes from a different operation---where gzip asks \textit{``what byte
sequences repeat?''}, PCAP-LM asks \textit{``what information does an analyst
actually need?''}, discarding Ethernet framing, IP headers, exact sequence
numbers, sub-millisecond timestamps, and payload bytes, and re-encoding the
remainder in a domain-specific alphabet. PCAP-LM is to a raw PCAP as a radiology
report is to a raw MRI scan: it is a \emph{knowledge extraction} step, not an
entropy coder. The practical consequence for LLM applications is that a lossless
compressor delivers an opaque byte stream, whereas PCAP-LM delivers a readable
document with the right abstractions precomputed---anomalies surfaced in the
annex, size distributions rendered as sparklines, TLS session parameters on a
single line---that an LLM can parse directly with no decompression step.

\subsection{Lossy-but-Recoverable Design}

PCAP-LM is lossy by design: exact TCP sequence numbers, sub-millisecond
timestamp precision, and payload bytes are not preserved in the main document.
However, the \code{@REFS} side-index (emitted by \code{pcap2lm convert --refs})
maps every flow to its global PCAP frame numbers, enabling lossless drill-down:
an LLM analysis can cite \code{f1\#142} and a tool call can retrieve the raw
bytes from the original PCAP via \code{tshark -Y "frame.number~==~843"}. This
makes PCAP-LM suitable for investigative workflows where the analyst starts with
the compressed summary and expands specific packets on demand.

Because production captures embed identifying information (host IPs, SNI and DNS
names, capture filenames encoding node names), the recommended deployment
pseudonymises the document before submission to a cloud LLM and de-anonymises
the response afterwards. \code{pcap2lm anonymize} substitutes IPs, domains, and
filenames into reserved documentation ranges (RFC\,5737, RFC\,3849, RFC\,2606)
while leaving forensically load-bearing fields---JA3 fingerprints, sparklines,
glyph streams, counts, ports, \code{@LEGEND}, \code{@REFS}---untouched, and
writes the reverse mapping to a local file that never leaves the analyst's
environment.

\subsection{Limitations}

\textbf{Corpus homogeneity and BPE saturation.} Our evaluation corpus consists
exclusively of HTTPS bulk-download tests between fixed endpoints. BPE training
consequently exhausted at 140 merges (final vocabulary: 159 tokens): the model
is maximally efficient for this traffic type, but will not generalise to
heterogeneous environments (mixed enterprise protocols such as DNS, SMTP,
HTTP/2) without retraining, likely requiring 2\,000--4\,000 composite tokens.
Beyond vocabulary growth, three pressures compound: the glyph alphabet must grow
to encode protocol-specific state (QUIC stream boundaries, SMTP command phases,
HTTP/2 frame types); cross-protocol queries involve correlations an in-context
legend cannot convey; and \code{@LEGEND} overhead grows with each protocol
family. Under those conditions fine-tuning (Strategy~C,
\S\ref{sec:strategies}) becomes substantially more attractive, since it
eliminates the legend overhead, internalises the composite vocabulary, and
supports cross-protocol reasoning without additional scaffolding. We estimate
the threshold at $\geq$4 distinct protocol families, a stable glyph alphabet,
and a sufficiently diverse labelled (PCAP-LM, question, answer) dataset; until
then Strategy~A remains the lower-cost, more maintainable choice.

\textbf{Skewed baseline comparison.} The 51.0\% baseline accuracy is heavily
skewed by truncation: counting packets, measuring duration, and detecting late
retransmissions failed because the model was denied the data, not because
\code{tshark} represents it poorly. The comparison establishes that PCAP-LM
solves the context-window bottleneck, but it penalises the baseline for its
size rather than for any representational deficiency.

\textbf{Algorithmic blind spots.} Discarding exact sequence numbers and
sub-millisecond timestamps blinds PCAP-LM to certain nuances---most notably a
24\% false-negative rate for TCP retransmissions, because the simplified
single-pass duplicate-sequence detector fails when an upstream TCP proxy
modifies sequence numbers between sender and receiver.

\textbf{Untested anomaly heuristics.} The asymmetric-traffic heuristic
($>$100$\times$ byte ratio) was never exercised on real data: our bulk-download
corpus never triggers the threshold, so its efficacy remains verified only on
synthetic captures.

\textbf{Extensions.} Planned work includes delta-encoded delays (emitting the
delay glyph only when it changes, a further 10--20\% reduction), ACK-run
collapsing (5--15\%), fine-tuning on PCAP-LM documents, and an MCP companion
server exposing \code{expand\_flow}, \code{get\_packet}, and \code{search} for
interactive drill-down over one or many captures.

\section{Conclusion}
\label{sec:conclusion}

We have presented PCAP-LM, a flow-centric, four-layer LLM-native text format
that acts as a lossy \emph{knowledge extraction} step for network captures,
together with the PacketGlyphs encoding alphabet, a constrained PMI-BPE
tokenizer, and a motif run-length encoder. On a 60-file held-out test set of
5G/4G HTTPS production captures, the pipeline achieves an \textbf{812$\times$}
token reduction over \code{tshark~-V}, fitting the largest captures within a
single LLM context window. In a forensic question-answering evaluation over 30
files, a frontier LLM achieves \textbf{99.3\%} accuracy from PCAP-LM documents
versus 51.0\% from the token-budget-matched \code{tshark~-V} prefix,
demonstrating that PCAP-LM's semantic richness---flow topology, TLS metadata,
anomaly annotations, and behavioural patterns in plain text---enables LLM
analysis that verbatim packet transcripts cannot support at any feasible context
size.

This compression is inherently lossy: exact sequence numbers and raw payload
bytes are discarded, introducing blind spots such as a 24\% false-negative rate
for TCP retransmissions. The \code{@REFS} side-index mitigates this by mapping
every summarised flow back to its original PCAP frame numbers, enabling lossless
drill-down via a companion MCP server---a capability whose full validation
remains future work. Our pipeline is currently optimised for homogeneous TLS
bulk-download traffic; extending PCAP-LM to heterogeneous mixed-protocol
enterprise environments will require expanding the BPE vocabulary and
potentially fine-tuning LLMs for cross-protocol reasoning. We believe PCAP-LM
represents a crucial step toward making network traffic analysis a first-class task for large language models.

\bibliography{ieeectl,refs}

@IEEEtranBSTCTL{IEEEexample:BSTcontrol,
  CTLuse_url               = "no",
  CTLuse_forced_etal       = "yes",
  CTLmax_names_forced_etal = "5",
  CTLnames_show_etal       = "1",
  CTLuse_alt_spacing       = "yes",
  CTLdash_repeated_names   = "no"
}

@inproceedings{sennrich2016,
  author    = {Sennrich, Rico and Haddow, Barry and Birch, Alexandra},
  title     = {Neural Machine Translation of Rare Words with Subword Units},
  booktitle = {Proceedings of the 54th Annual Meeting of the Association
               for Computational Linguistics (ACL 2016)},
  pages     = {1715--1725},
  year      = {2016},
  address   = {Berlin, Germany},
  url       = {https://aclanthology.org/P16-1162/},
  note      = {arXiv:1508.07909}
}

@misc{netgpt2023,
  author        = {Meng, Xuying and Lin, Chungang and Wang, Yequan and Zhang, Yujun},
  title         = {{NetGPT}: Generative Pretrained Transformer for Network Traffic},
  year          = {2023},
  eprint        = {2304.09513},
  archivePrefix = {arXiv},
  primaryClass  = {cs.NI},
  url           = {https://arxiv.org/abs/2304.09513}
}

@inproceedings{etbert2022,
  author    = {Lin, Xinjie and Xiong, Gang and Gou, Gaopeng and Li, Zhen
               and Shi, Junzheng and Yu, Jing},
  title     = {{ET-BERT}: A Contextualized Datagram Representation with
               Pre-training Transformers for Encrypted Traffic Classification},
  booktitle = {Proceedings of the ACM Web Conference 2022 (WWW '22)},
  pages     = {633--642},
  year      = {2022},
  address   = {Lyon, France},
  doi       = {10.1145/3485447.3512217},
  note      = {arXiv:2202.06335}
}

@inproceedings{llmcap2024,
  author    = {Tulczyjew, L. and Jarrah, K. and Abondo, C.
               and Bennett, D. and Weill, N.},
  title     = {{LLMcap}: Large Language Model for Unsupervised {PCAP}
               Failure Detection},
  booktitle = {IEEE International Conference on Communications (ICC)
               Workshop on the Impact of Large Language Models on 6G Networks},
  year      = {2024},
  note      = {arXiv:2407.06085}
}

@misc{trafficllm2025,
  author        = {Cui, Tianyu and Lin, Xinjie and Li, Sijia and Chen, Miao
                   and Yin, Qilei and Li, Qi and Xu, Ke},
  title         = {{TrafficLLM}: Enhancing Large Language Models for Network
                   Traffic Analysis with Generic Traffic Representation},
  year          = {2025},
  eprint        = {2504.04222},
  archivePrefix = {arXiv},
  primaryClass  = {cs.CR},
  url           = {https://arxiv.org/abs/2504.04222}
}

@misc{ja32017,
  author       = {Althouse, John and Atkinson, Jeff and Atkins, Josh},
  title        = {{JA3}: {SSL/TLS} Client Fingerprinting for Malware
                  Detection and Hunting},
  year         = {2017},
  howpublished = {Salesforce Engineering / GitHub},
  url          = {https://github.com/salesforce/ja3}
}

@misc{scapy,
  author       = {Biondi, Philippe},
  title        = {Scapy: Interactive Packet Manipulation Program},
  year         = {2003},
  howpublished = {Presented at LSM 2003},
  url          = {https://scapy.net}
}

@misc{wireshark,
  author       = {{Wireshark Foundation}},
  title        = {Wireshark Network Analyser},
  year         = {2025},
  howpublished = {\url{https://www.wireshark.org}},
  note         = {Version~4.x; originally released 1998}
}

@misc{anthropic2025,
  author       = {{Anthropic}},
  title        = {Claude Sonnet~4.6},
  year         = {2025},
  howpublished = {\url{https://www.anthropic.com}}
}

@misc{libpcap,
  author       = {Jacobson, Van and Leres, Craig and McCanne, Steven},
  title        = {libpcap: Packet Capture Library},
  year         = {1994},
  howpublished = {Lawrence Berkeley National Laboratory},
  url          = {https://www.tcpdump.org}
}

\end{document}